\documentclass[10pt,  twocolumn, twoside, letterpaper, conference, final]{IEEEtran}
\IEEEoverridecommandlockouts
\usepackage{cite}
\usepackage{algorithmicx}
\usepackage[ruled]{algorithm}
\usepackage{algpseudocode}
\usepackage{mathtools}
\usepackage{dsfont}

\usepackage{amsmath}
\usepackage{amsfonts}
\usepackage{amssymb}

\usepackage{subcaption}
\usepackage{multirow}
\usepackage{textcomp}
\def\BibTeX{{\rm B\kern-.05em{\sc i\kern-.025em b}\kern-.08em
    T\kern-.1667em\lower.7ex\hbox{E}\kern-.125emX}}
\usepackage{color}
\usepackage[keeplastbox]{flushend}
\usepackage{enumerate}
\usepackage{mathrsfs}
\usepackage{mathtools}
   \usepackage{graphics}

\usepackage{amsmath,amsfonts,amssymb}

\usepackage{url}

\usepackage{hyperref}

\newcommand{\x}{\mathbf{x}}
\newcommand{\z}{\mathbf{z}}
\newcommand{\B}{\mathbf{B}}
\newcommand{\W}{\mathbf{W}}

\begin{document}

\title{Deep Learning Based Detection for Spectrally Efficient FDM Systems}

\author{David Picard,~\IEEEmembership{Member,~IEEE,}
        Arsenia Chorti,~\IEEEmembership{Senior Member IEEE}}

\maketitle

\begin{abstract}
In this study we present how to approach the problem of building efficient detectors for spectrally efficient frequency division multiplexing (SEFDM) systems. The superiority of residual convolution neural networks (CNNs) for these types of problems is demonstrated through experimentation with many different types of architectures.
\end{abstract}



\section{Introduction}
In this report we present how to approach the problem of building efficient detectors for spectrally efficient frequency division multiplexing (SEFDM) systems \cite{1}, \cite{2}, \cite{3}, \cite{4}, \cite{5}. We provide numerical results for standard deep networks and demonstrate that the insertion of residual connections is important die to vanishing eigenvalues in the system Gram matrix. Finally, the superiority of residual convolution neural networks (CNNs) for these types of problems is demonstrated after experimentation with many different types of architectures. These results exhibit the optimization steps taken during the final selection presented in \cite{6}.

\section{Problem statement}

We transmit symbols $z \in \mathbb{C}$ in packets on $N$ using frequency division multiplexing. The symbols corresponds to $M$ possible different classes called $c(z)$  (\textit{e.g.} 4 in case of qpsk). To that end, $N$ symbols are stacked in a vector $\z \in \mathbb{C}^N$ and projected into frequency bins using the projection matrix $\B \in \mathbb{C}^{N\times N}$. During the transmission, a noise $\epsilon \sim \mathcal{N}(0, \sigma)$ is added, which leads to the observed vector $\x \in \mathbb{C}^N$:

\begin{align}
    \x = \B^\dag(\B \z + \epsilon)
\end{align}

In OFDM, $\B$ is orthonormal, hence $\B^\dag\B = I$ and recovering the classes of the emitted symbols is \emph{easy} since the optimal detector on $x$ corresponds to the Voronoï partition on $z$ (equivalent to MLE).

In case $\B$ is no longer orthonormal (overlapping frequency bins), the Voronoï partition on $z$ is deformed in a non trivial way. It is then no longer easy to recover $c(z)$ from the corresponding $x$. We thus want to learn a prediction function $f$ whose goal is to predict the class $c(z)$ of the emitted symbol $z$ given the corresponding received symbol $x$:

\begin{align}
    \min _f \sum_z \mathbb{E}_\epsilon[ l(f(x), c(z))]
\end{align}

With $l(\cdot, \cdot)$ being a cost function measuring the error between $f(x)$ and $c(z)$. The optimal $f$ is obtained by performing a gradient descent over $l$.

\section{Neural architectures}

We propose to implement $f$ using a deep neural network. Here, a neural network is a composition of parametric functions $h_l: \mathbb{R}^{2N} \rightarrow \mathbb{R}^{w}$ (instead of complex numbers, we manipulate 2 dimensional real vectors), followed by a projection $g: \mathbb{R}^D \rightarrow \mathbb{R}^M$ onto the classes simplex.

Each function is called a \emph{layer}. A neural network is characterized by 3 parameters: the family of functions $h$ that are used, its depth $d$ which corresponds to the number of layers and its width $w$ which corresponds to the size of the intermediate space which the inputs are mapped to. Note that since $h$ are not linear, it makes sense to increase the dimension ($w \geq 2N$) to \emph{unfold} the transformation that was performed by $\B$.
The gradient descent is performed over all parameters of all layers using the chain rule.

We now described the different families of layers that are used in this study.

\subsection{Multiple Layer Perceptron}

Multiple Layer Perceptron (MLP) is the historic neural networks. A layer is composed of a linear projection followed by a non-linear element-wise activation function:

\begin{align}
    \x_l = h_l(\x_{l-1}) = s(\W_l\x_{l-1})
\end{align}

With $\W_l \in \mathbb{R}^{[2N, w]\times w}$ are the weights of the layers and $s(t) = \max(0, t)$ is its activation function (simple rectification here).

\subsection{Residual MLP}
MLP are notoriously difficult to train because of exploding/vanishing gradients due to the increased depth (if $\lambda$ is a bound to the singular values of the weights, then the gradient on the first layers is in $\mathcal{O}(\lambda^d)$).

To mitigate this problem, residual connections can be added:

\begin{align}
    \x_l = h_l(\x_{l-1}) = s(\W_l\x_{l-1}) + \x_{l-1}
\end{align}

Usually, residual - or skip - connections are not added at every transform, but every 2:

\begin{align}
    \x_{1/2} = s(\W_l\x_{l-1})\\
    h_l(\x_{l-1}) =  \x_{l} = s(\W_l'\x_{1/2}) + \x_{l-1}
\end{align}

\subsection{Convolutional neural networks}

When the input signal is structured (\textit{e.g}, time series), it makes sense to apply an MLP layer on a sliding window. This corresponds to a convolution with the weight matrix, followed by the non linear activation:

\begin{align}
    \x_l = [s(\W_{il} \star \x_{l-1})]_i
\end{align}

With $\W_{il} \in \mathbb{R}^{k\times d}$, where $k$ corresponds to the window size (\emph{kernel size}). The output is a stack of several of such convolutions, hence the name Convolutional Neural Networks (CNN). 

If the input signal is not structured it makes sometimes nonetheless sense to use a CNN. Indeed, the exact same output could have been obtained with an MLP with carefully chosen weights. In such MLP, most of the weights are $0$ (corresponding to elements outside of the window). The non zero weights are duplicated since the weights are independent of the position of the window. Achieving an MLP with such structural constraint ($\ell_0$ norm, low diversity) is extremely difficult, while it exists by construction in CNN.

\section{Experiments}

\subsection{Architecture influence}

In these experiments, we use the following parameters:
\begin{itemize}
    \item $M = 4$ classes
    \item $N = 32$ sub-carriers
    \item $\alpha = 0.1$ overlap between frequency bins
\end{itemize}

\begin{figure}
     \centering
     \includegraphics[width=\columnwidth]{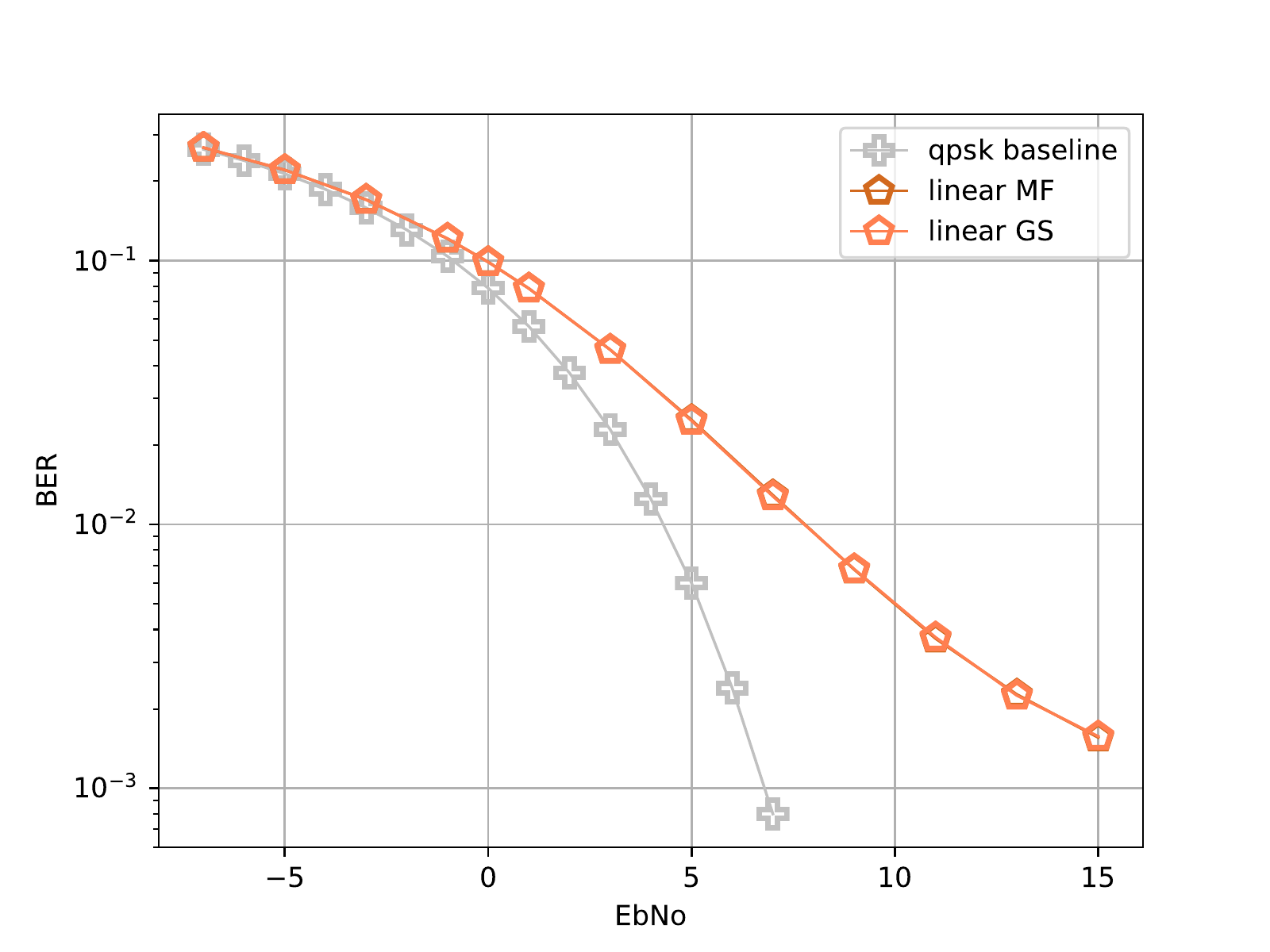}
     \caption{Baseline}
     \label{fig:baseline}
 \end{figure}

Our baseline comparisons are the ML for non multiplexed signal using QPSK and a trained linear predictor.
The linear predictor is able to obtain the exact same performances as the ML QPSK in the OFDM case.
In the non orthogonal case ($\alpha > 0$), it should be slightly better than simply detecting along the axes as no such constraint exists. It doesn't take into account the shape of the transformed noise and thus is not very satisfying.
Figure \ref{fig:baseline} shows the bit error rate of the baselines for different SNR.
Using the matching filter (MF) ofr the Gram-Schmidt orthogonalized filter (GS) does not change the performances of the linear predictor.

We train the model on $2.10^8$ randomly generated received symbols and we evaluate the bit error rate on $4.10^6$ symbols. BER below $10^{-5}$ are thus not significant.

 For each architecture, we explore depth $d$ and width $w$ to see the influence of each characteristic.
 
 \subsection{MLP}
 

     \begin{figure}
    \includegraphics[width=\columnwidth]{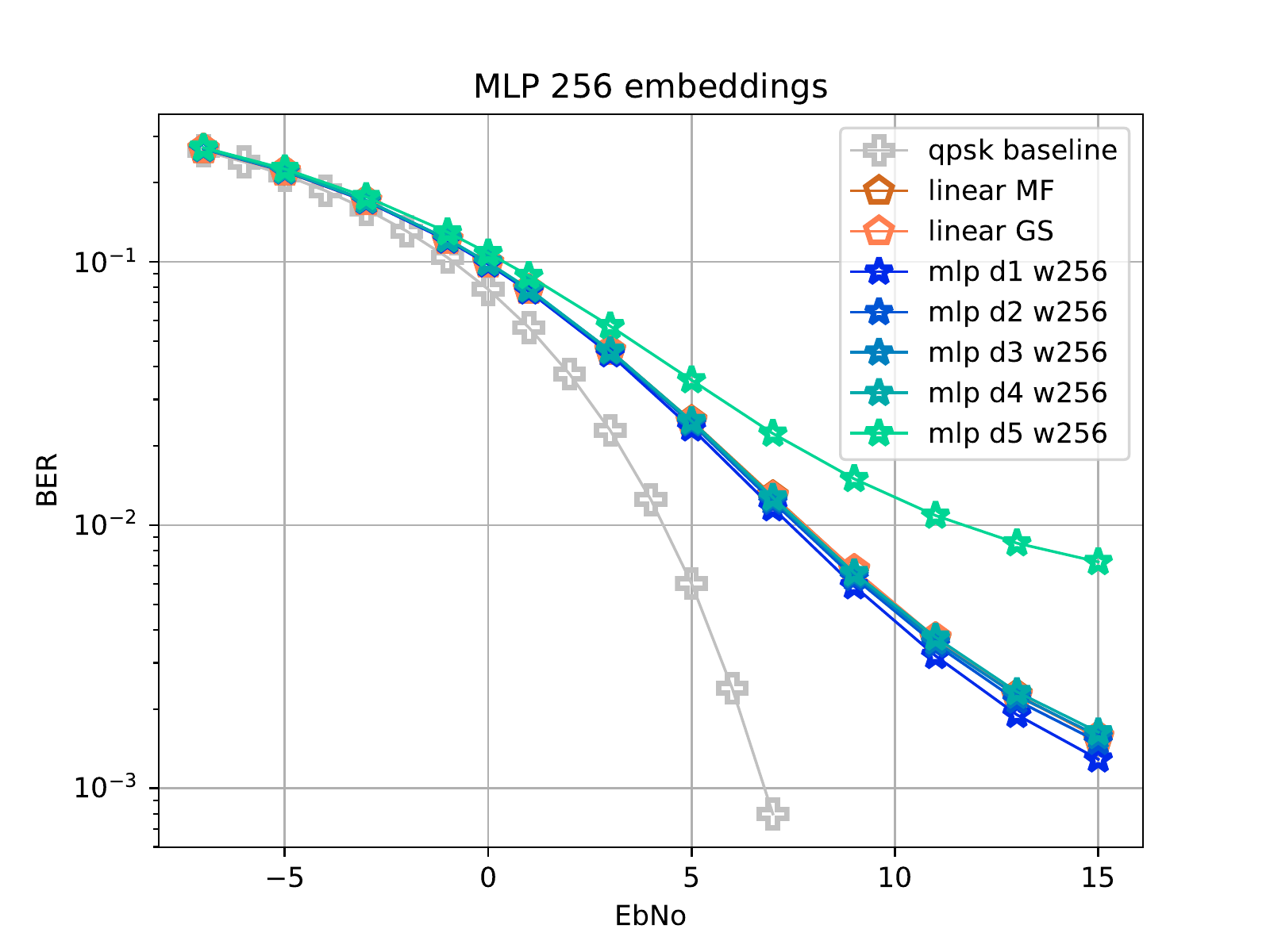}\label{fig:mlp256}
     \caption{MLP}
 \end{figure}
 
 Adding width improves the results only for shallow networks.
 
 \subsection{Residual MLP}

      

 \begin{figure}
     \centering
      \includegraphics[width=\columnwidth]{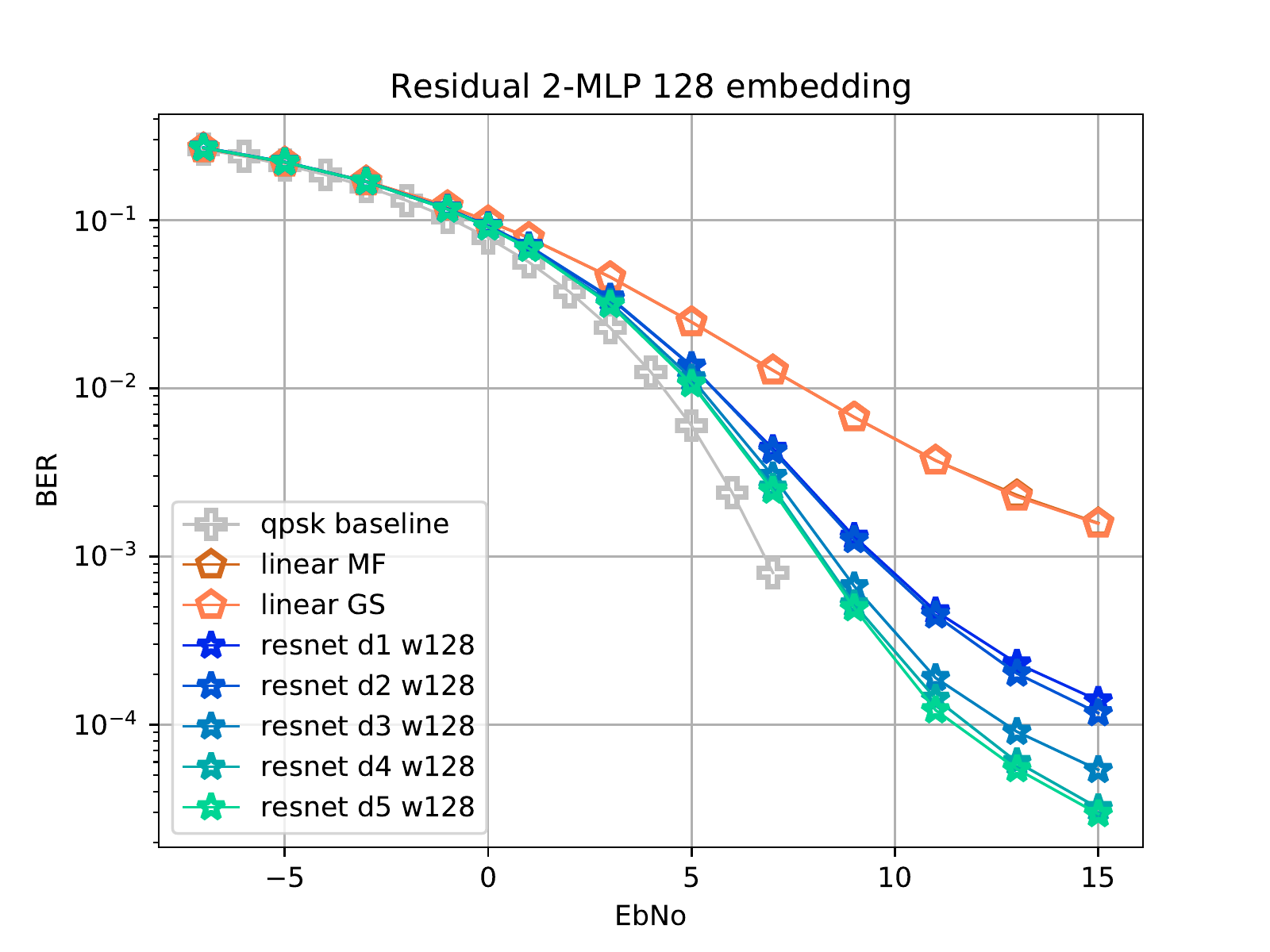}\label{fig:2res128}
      \caption{Residual 2-MLP}
      \end{figure}
      
      \begin{figure}
      \includegraphics[width=\columnwidth]{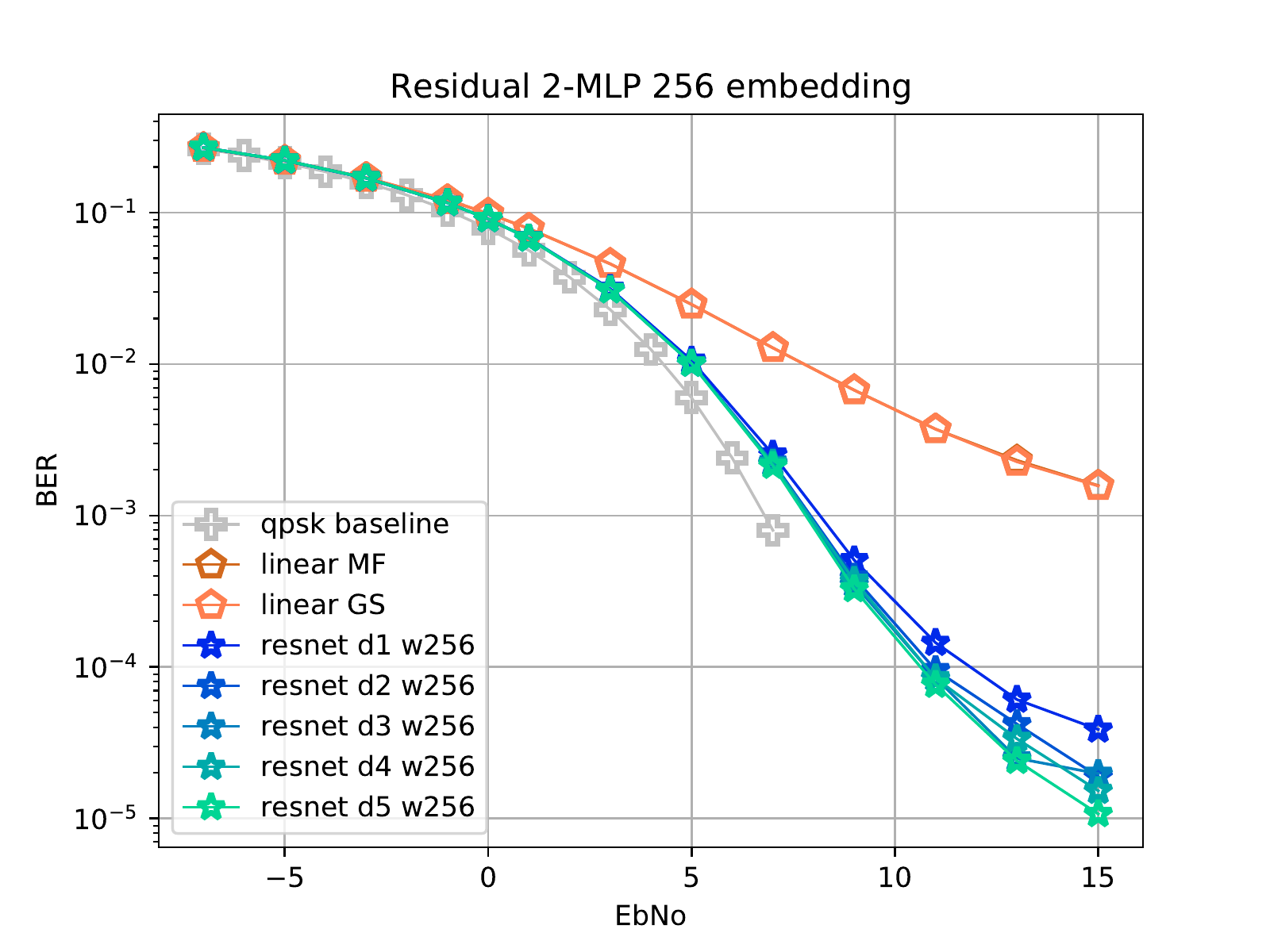}\label{fig:2res256}
     \caption{Residual 2-MLP}
 \end{figure}

Residual MLP are able to improve over the linear predictor, especially with the more common 2 block architecture between skip connections. Adding width tend to improve shallow networks only.

 \subsection{CNN}

 \begin{figure}
     \centering
     \includegraphics[width=\columnwidth]{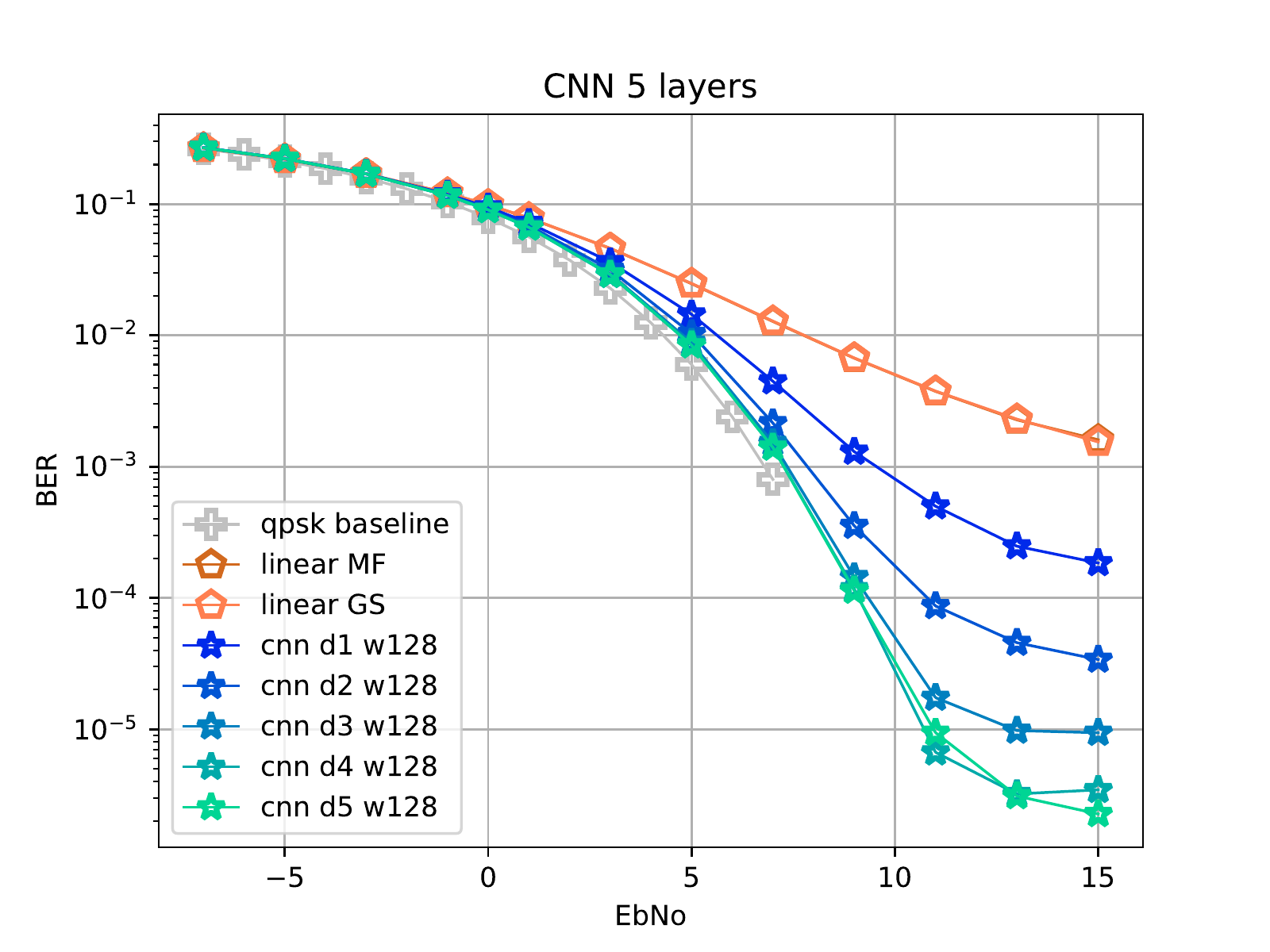}\label{fig:cnn128}
     \caption{CNN}
     \end{figure}
     
     \begin{figure}
    \includegraphics[width=\columnwidth]{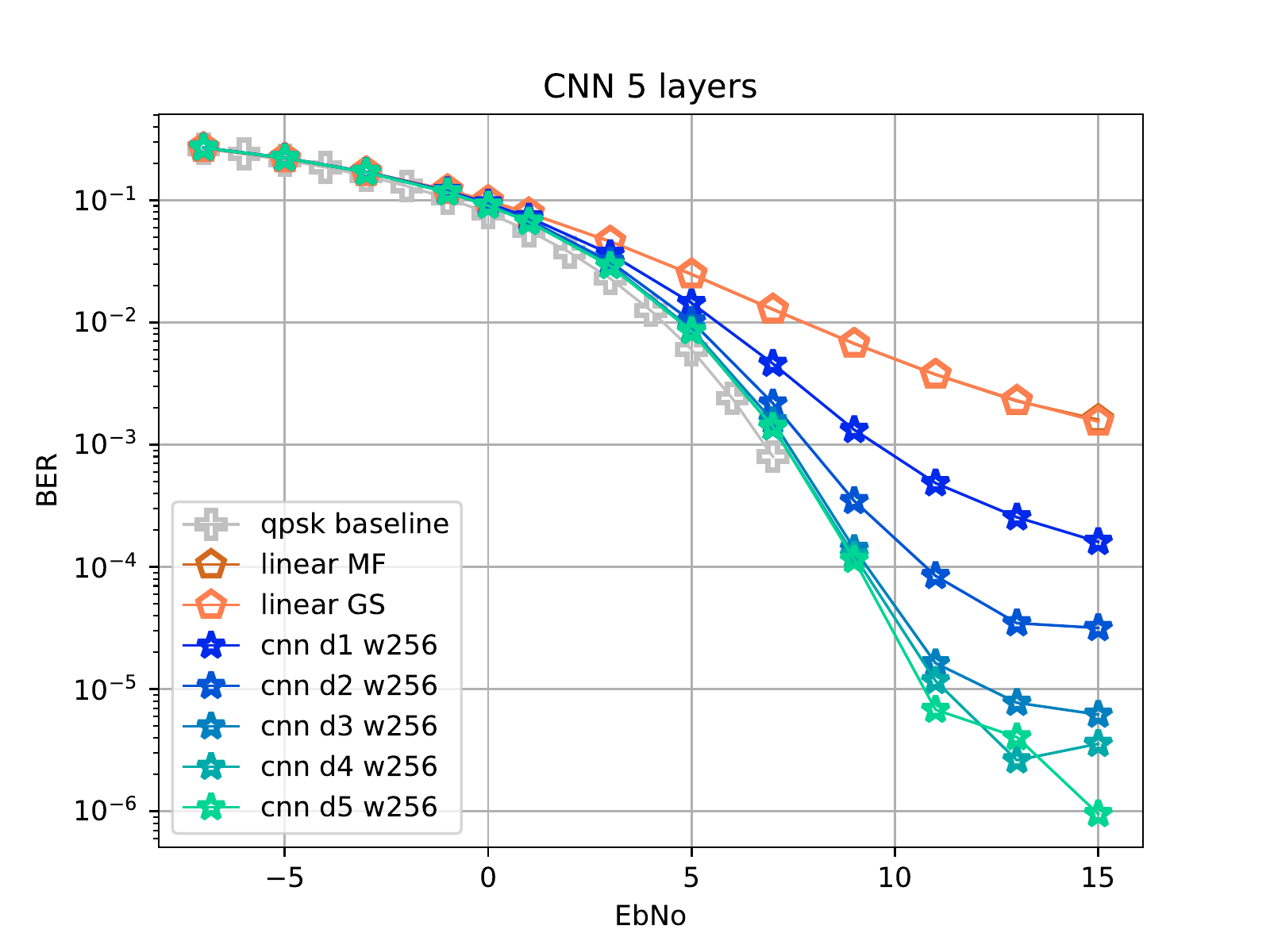}\label{fig:cnn256}
     \caption{CNN}
 \end{figure}
 
 CNNs, provide a major improvement and are about 1dB behind the orthogonal case. This shows that the structural constraints (sparsity, repeatability) on the predictor are making the learning problem easier, which leads to a better solution.
Contrarily to MLP, adding width does not significantly improve the results.

 \subsection{Residual CNN}
 
     
  
 \begin{figure}
     \centering
     \includegraphics[width=\columnwidth]{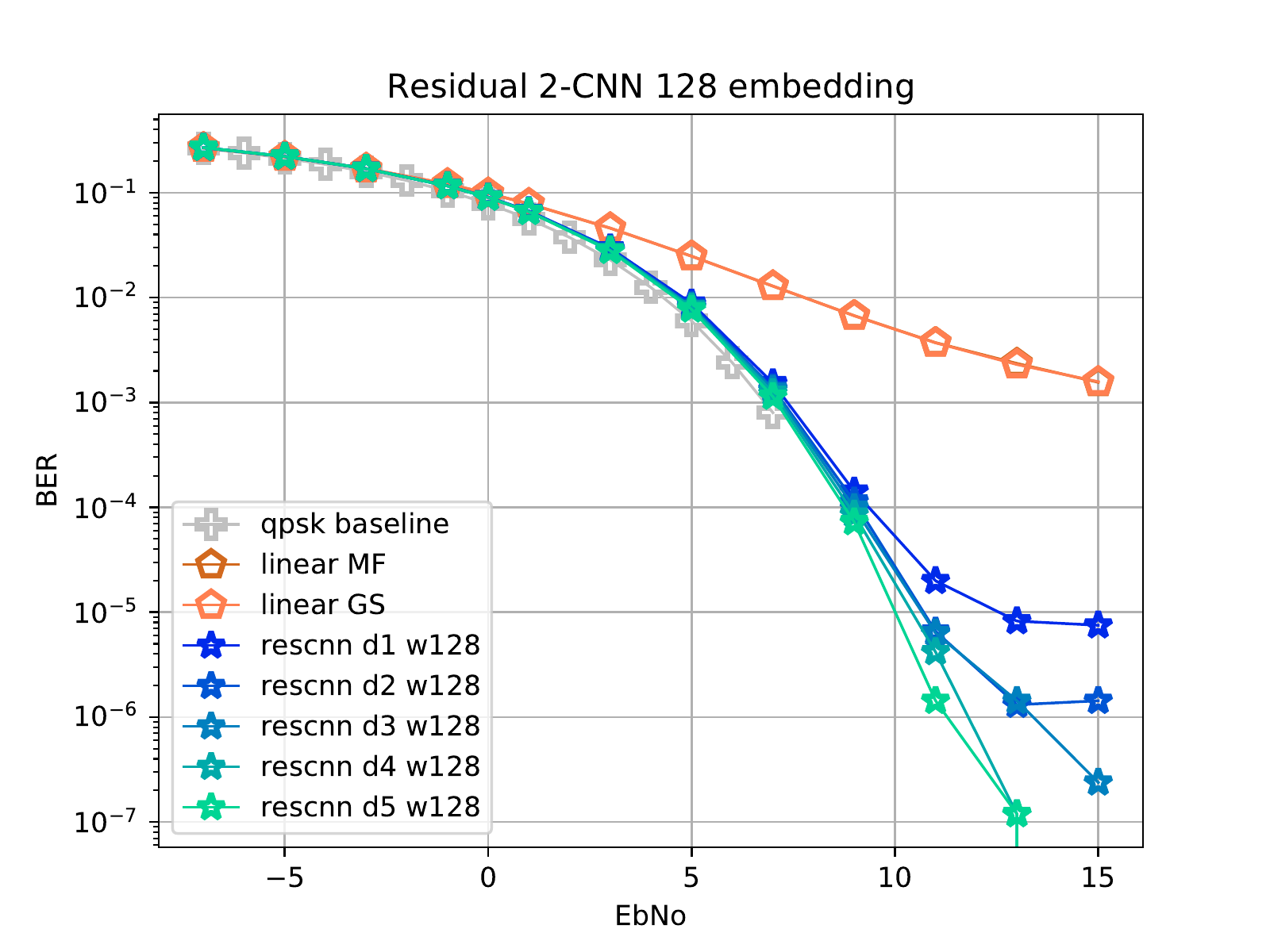}\label{fig:2rescnn128}
     \end{figure}
     
     \begin{figure}
     \includegraphics[width=\columnwidth]{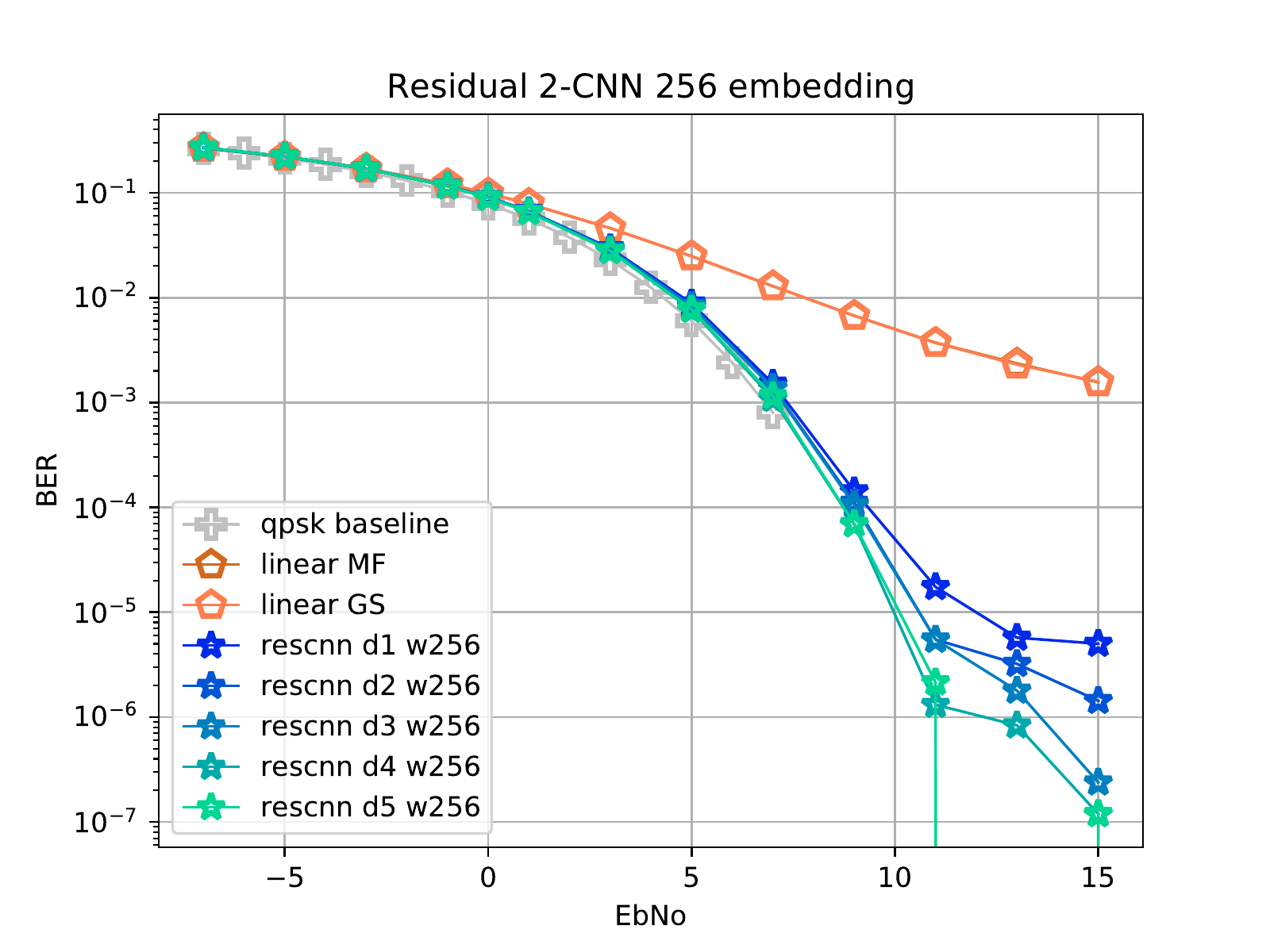}\label{fig:2rescnn256}
     \caption{Residual 2-CNN}
 \end{figure}

Residual CNN improve over the simple convolution, especially with the more common 2 block architecture between skip connections which is deeper. This shows that residual connection are beneficial to deeper architectures (which is why they were invented for), and that deeper architectures, when trained properly, can reached better performances. Adding width slightly improves the error rate ate an increased model size cost.

%
\IEEEpeerreviewmaketitle



\bibliographystyle{IEEEtran}
%
%









\end{document}